\documentclass[11pt,a4paper]{article}
\usepackage{flafter}
\usepackage[dvips]{graphicx}
\usepackage{latexsym}
\usepackage[fleqn]{amsmath}
\usepackage{algorithmic,algorithm,amsfonts,amssymb}

\newtheorem{definition}{Definition}

\begin{document}

\title{DDMF: An Efficient Decision Diagram Structure for Design Verification of Quantum Circuits under a Practical Restriction}

\author{
Shigeru Yamashita$^{1}$ 
$\quad$
Shin-ichi Minato$^{2}$ 
$\quad$
D. Michael Miller$^{3}$\\
\\
$^{1}$Graduate School of Information Science, Nara Institute of Science 
and Technology\\ 
{\tt ger@is.naist.jp}\\
$^{2}$Graduate School of Information Science and Technology, Hokkaido University\\
{\tt minato@st.hokudai.ac.jp}\\
$^{3}$Faculty of Engineering, University of Victoria\\
{\tt mmiller@cs.uvic.ca}
}
\date{}

\maketitle

\begin{abstract}
Recently much attention has been paid to quantum circuit
design to prepare for the future ``quantum computation era.'' Like
the conventional logic synthesis, it should be important to verify and
analyze the functionalities of generated quantum circuits. For that
purpose, 
we propose an efficient verification method for quantum circuits under
a practical restriction. Thanks to the restriction, we can introduce
an efficient verification scheme based on decision diagrams called
{\it Decision Diagrams for Matrix Functions (DDMFs)}. Then, we show
analytically the advantages of our approach based on DDMFs
over the previous verification techniques. In order to introduce DDMFs, we also  
introduce new concepts, {\it quantum functions} and {\it matrix functions}, which
may also be interesting and useful on their own for designing quantum
circuits.
\end{abstract}

\hspace{1cm}{\bf keywords:~}
quantum circuit, verification, decision diagram

\section{Introduction}


Recently {\it quantum computing} has attracted great attention by its
potential abilities~\cite{Nielsen2000}. 
To realize a quantum algorithm, it is necessary to design the 
 corresponding {\it quantum circuit} as small as possible. 
Thus, it should be very important to study quantum circuit design
methods even before quantum computing is physically realized. 
Indeed, there has been a great deal of
research~\cite{ger2002-1,SPMH03,MDM05,AP06,MYM05,HSYYP04,SMB06,MVBS04}
for 
quantum circuit design.  

Typical quantum circuit design methods  are based on
{\it matrix decomposition}~\cite{SMB06,MVBS04} since a quantum
algorithm is expressed by a matrix. They can treat any kind of quantum
circuits, but they cannot treat large (hence, practical) size problems 
since they need to express  matrices explicitly and thus they need
exponential time and memory. (Note that a matrix for an $n$-bit
quantum circuit is $2^n \times 2^n$, which will be explained later.)   

There is a different approach for quantum circuit
design~\cite{ger2002-1,SPMH03,MDM05,AP06,MYM05,HSYYP04}. 
The approach is to focus on quantum circuits calculating only (classical) 
Boolean functions by the following observation~\cite{ger2002-1}: 
Standard quantum algorithms usually consist of two parts, which we call
{\it common parts} and {\it unique parts} below. {\it Common parts} do
not differ for each problem instance.  On the other hand, {\it unique
 parts} differ for each problem instance. For example, Grover search
algorithm~\cite{Gro96}, one of the famous 
quantum algorithms, consists of so called an {\it oracle} part and the other 
part. An {\it oracle} part calculates (classical) 
Boolean functions depending on the specification of a given problem 
instance, while the other part consists of some quantum specific 
operations and does not change for all the problem instances.  
When we developed a new quantum algorithm, we should have designed the
common part. Therefore, we do not need to design the common part for
individual problem instances. On the other hand, since the unique part
of a quantum algorithm differ for each problem instance, we need to
have efficient design and verification methods for that part.  Since unique part 
calculates classical Boolean functions,  by focusing on only 
unique parts, we may have a design method to handle practical size
problems based on (classical) logic  synthesis techniques,
especially reversible logic synthesis techniques. Indeed 
there has been a great deal of research focusing on quantum circuits 
to calculate 
classical Boolean functions in the conventional logic synthesis  research
 community~\cite{ger2002-1,SPMH03,MDM05,AP06,MYM05,HSYYP04}.  
We also focus on this type of quantum circuits in this paper. 
It should be noted that there are many different points between our
target quantum circuits and the conventional logic circuits (as will be
explained later) although our target quantum circuits calculate only
classical Boolean functions. This is because we need to implement
circuits with quantum specific operations (as will be explained later). 
Therefore, we definitely need quantum specific design and verification
methods even for our target quantum circuits. 


Recently a paper~\cite{VMH07} discussed a problem of the equivalence check
of {\it general} quantum circuits and quantum states considering 
the so-called 
{\it phase equivalence} property of quantum states. Even for quantum
circuits calculating {\it only} Boolean functions, it should be very 
important to verify and analyze the functionalities of designed circuits
as in the case of classical logic synthesis. 
For example, we may consider the following situation: 
One of the possible realizations of quantum computation is considered 
to be so called a {\it linear-nearest-neighbor (LNN)} architecture 
in which the {\it quantum bits (qubits)} are arranged on a line, and
only operations to neighboring qubits are allowed. Thus, we need
to modify a designed quantum circuit so that it uses only gates that 
operate to two adjacent qubits. In such a case, we may use some
complicated transformations by hand, and thus it is very convenient if we have a verification
tool to confirm that the original and the modified quantum circuits 
are functionally equivalent. 

If we consider only the classical type gates, it is enough to use the
conventional verification technique such as Binary Decision Diagrams
(BDDs)~\cite{brya86} for the verification.    
However, even if we consider quantum circuits calculating only Boolean 
functions, it is known that non-classical (quantum specific) gates are
useful to reduce the circuit size~\cite{AP06,MYM05,HSYYP04,cuccaro-2004}. Thus we
need to verify quantum circuits with non-classical gates. In such
cases, a classical technique is obviously not enough. 

As for simulating quantum circuits, 
 efficient techniques using decision diagrams 
such as Quantum Information Decision Diagrams (QuIDDs)~\cite{VMH05}
and Quantum Multiple-valued Decision Diagrams (QMDDs) ~\cite{MT06} 
 have been proposed. 
By using these efficient diagrams, we can express the functionalities of
two quantum circuits, and then verify the equivalence of the two 
circuits. However, they are originally proposed to simulate {\it 
  general} quantum circuits, and thus there may be a more efficient
method that is suitable for verifying the functionalities of quantum
circuits only for Boolean functions. 

{\bf Our contribution described in this paper.} 
Considering the above discussion, we introduce a new quantum circuit 
class: {\it Semi-Classical Quantum Circuits
  (SCQCs)}. 
Although SCQCs have a restriction, the class of SCQCs
covers  
all the quantum circuits (for calculating a Boolean function) 
 designed by the existing methods~\cite{HSYYP04}. 
Moreover, because of the 
restriction of SCQCs, we can express the  functionalities of SCQCs very 
efficiently as in the case of conventional verifications by BDDs.  For that 
purpose, we  introduce a new decision diagram structure called {\it a
  Decision Diagram for a  Matrix Function (DDMF)}. Then, we show that
the verification method based on DDMFs are 
much more efficient than the above mentioned methods based on
previously known techniques. 
We provide an analytical comparison between DDMFs and QuIDDs, and
reveal the essential difference: (1) We show that their  
ability to express the functionality of one quantum gate is 
essentially the same, but (2) we also show that our approach based on
DDMFs is  much more efficient for the verification of SCQCs than a method 
based on QuIDDs. (Note that this does not mean that
DDMFs are better than QuIDDs: DDMFs are only for SCQCs, whereas QuIDDs  
can treat all kinds of quantum circuits.) 
Moreover, we show by preliminary experiments that DDMFs can be used to
verify SCQCs of practical size (60 inputs and 400 gates). 
In order to introduce DDMFs, we also introduce new concepts, {\it
 quantum functions} and {\it matrix functions}, which may be 
interesting and useful on their own for designing quantum circuits
with quantum specific gates. 

\section{Semi-Classical Quantum Circuits and Their Representations by
  Decision Diagrams}\label{sec:basic}

This section introduces new concepts: SCQCs together with 
quantum functions, matrix functions and DDMFs.   

\subsection{Quantum States and Quantum Gates}\label{subsec:basic}

Before introducing our new concepts, let us briefly explain the basics
of quantum computation.

\begin{figure}[tb]
\begin{center}
\begin{minipage}{8cm}
 \resizebox{!}{2cm}{
 \includegraphics*{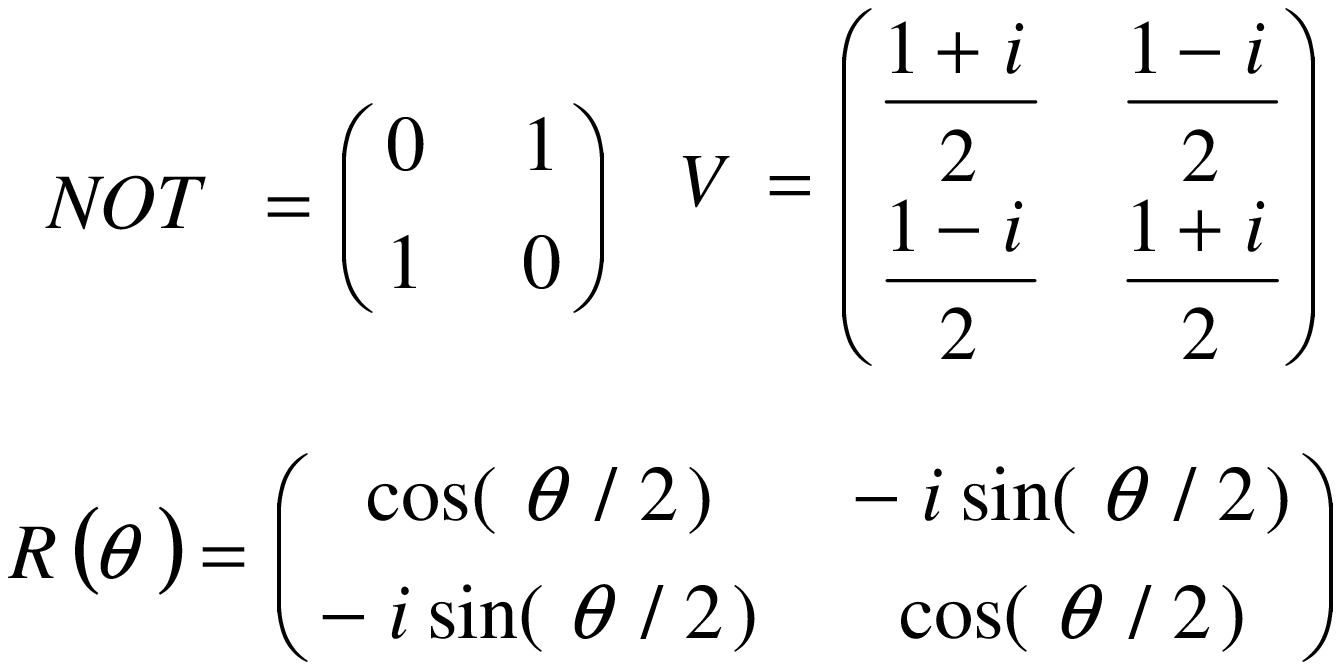}
 }
\caption{Unitary matrices}\label{fig:matrix}
\end{minipage}
\begin{minipage}{8cm}
\begin{center}
\resizebox{!}{2cm}{
\includegraphics*{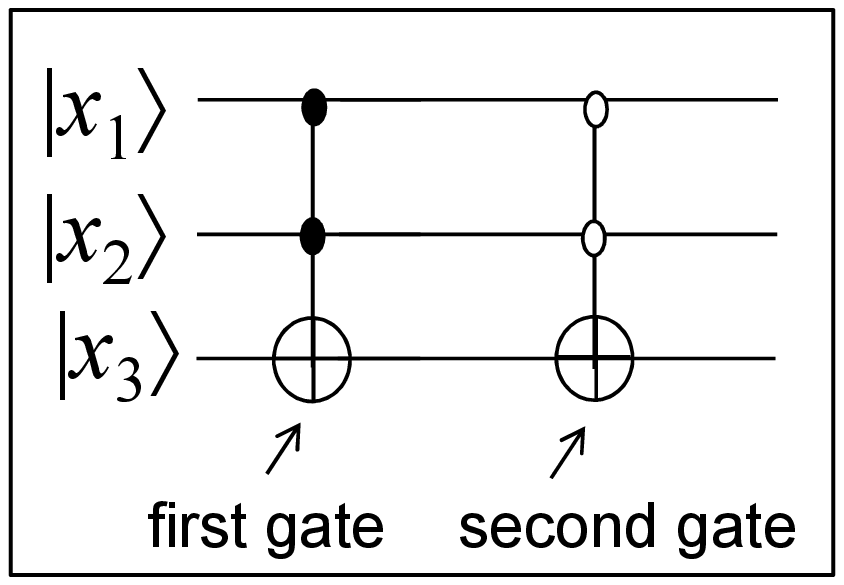}
}
\caption{A quantum circuit }\label{fig:Circuit1}
\end{center}
\end{minipage}
\end{center}

\end{figure}

In quantum computation, it is assumed that we can use a {\it
qubit} which is an abstract model of a {\it quantum state.} A
qubit can be described as
$\alpha\left|0\right>+\beta\left|1\right>$, where $\left|0\right>$
and $\left|1\right>$ are two basic states, and $\alpha$ and
$\beta$ are complex numbers such that $|\alpha|^2 + |\beta|^2 =
1$. It is convenient to use the following vectors to denote
$\left|0\right>$ and $\left|1\right>$, respectively:
$\left|0\right> =
  \left(
         \begin{matrix}
         1  \\
         0  \\
         \end{matrix}
\right)
$
and,
$
 \left|1\right> =
  \left(
         \begin{matrix}
         0  \\
         1  \\
         \end{matrix}
 \right).
$
Thus, $\alpha\left|0\right>+\beta\left|1\right>$ can be described as a
vector: 
$
 \alpha\left|0\right>+\beta\left|1\right> =
 \left(
         \begin{matrix}
         \alpha  \\
         \beta  \\
         \end{matrix}
 \right).
$
Then, any quantum operation on a qubit can be described as
a 2$\times$2 matrix. By the laws of quantum mechanics, the matrix must be
{\it unitary.} We call such a quantum operation a {\it quantum gate}.
For example, the operation which transforms $\left|0\right>$ and
$\left|1\right>$ to $\left|1\right>$ and $\left|0\right>$,
respectively, is called a $NOT$ gate whose matrix representation
is as shown in Fig.~\ref{fig:matrix}.

In addition to the above $NOT$ gates, we can also use any quantum
specific unitary matrix in quantum circuits. 
For example, {\it  rotation gates} denoted by $R(\theta)$ are often
used in quantum computation. The matrix for the gates is as shown in
Fig.~\ref{fig:matrix}. Although the functionality of rotation gates is
not classical, they are useful to design quantum circuits even for
(classical) Boolean functions~\cite{AP06}. 
Another quantum specific gate called  $V$ gate is  
also utilized to design quantum circuits for Boolean
functions~\cite{MYM05,HSYYP04}. The matrix for the gate is as shown
in Fig.~\ref{fig:matrix}. 
This gate has the interesting property that $V^2 = NOT$.



In the following, our primitive gates are (generalized) {\it controlled-U gates} 
which are defined as follows: 

\begin{definition}
A controlled-U gate has (possibly many) positive and negative control bits, and
one target bit. 
It applies a 2$\times$2 unitary matrix $U$ to the target qubit when 
the states of all the positive control bits are the states $\left|1\right>$ and 
the states of  all the negative control bits are  the state $\left|0\right>$.  
A controlled-U gate may not have a control bit. In such a case, it always 
applies $U$ to the target qubit. 
\end{definition}


See an example of a quantum circuit consisting of two controlled-$NOT$
gates in Fig.~\ref{fig:Circuit1}. 
This circuit has three qubits, 
$\left|x_1\right>$, $\left|x_2\right>$ and $\left|x_3\right>$, each
of which corresponds to one line. 
In quantum circuits, each gate 
works one by one from the left to the right. For the first gate, 
the target bit is $x_3$ and the symbol $\oplus$ means the 
 $NOT$ operation. The positive control bits are $x_1$ and $x_2$
 denoted by black  circles. This gate  performs $NOT$ on
 $\left|x_3\right>$ only when both $\left|x_1\right>$ and 
 $\left|x_2\right>$ are the  state $\left|1\right>$. Consider the second gate in
 the  same figure. The white circles denote negative controls, which
 means the gate performs $NOT$ only when both 
 $\left|x_1\right>$ and $\left|x_2\right>$ are the states $\left|0\right>$. 

In addition to controlled-$NOT$ gates which are essentially classical gates,
we can consider any (quantum specific) unitary operation for
controlled gates. For  
example, the functionalities of controlled gates in Figs.~\ref{fig:adder}
and ~\ref{fig:NON-SCQC} are various (e.g., NOT, $V$, $V^{-1}$,
$R(\frac{1}{2}\pi)$ and $R(\frac{1}{4}\pi)$). 

\subsection{Semi-Classical Quantum Circuit (SCQC)}

\begin{figure}[tb]
\begin{minipage}{10cm}
\begin{center}
\resizebox{!}{1.5cm}{
\includegraphics*{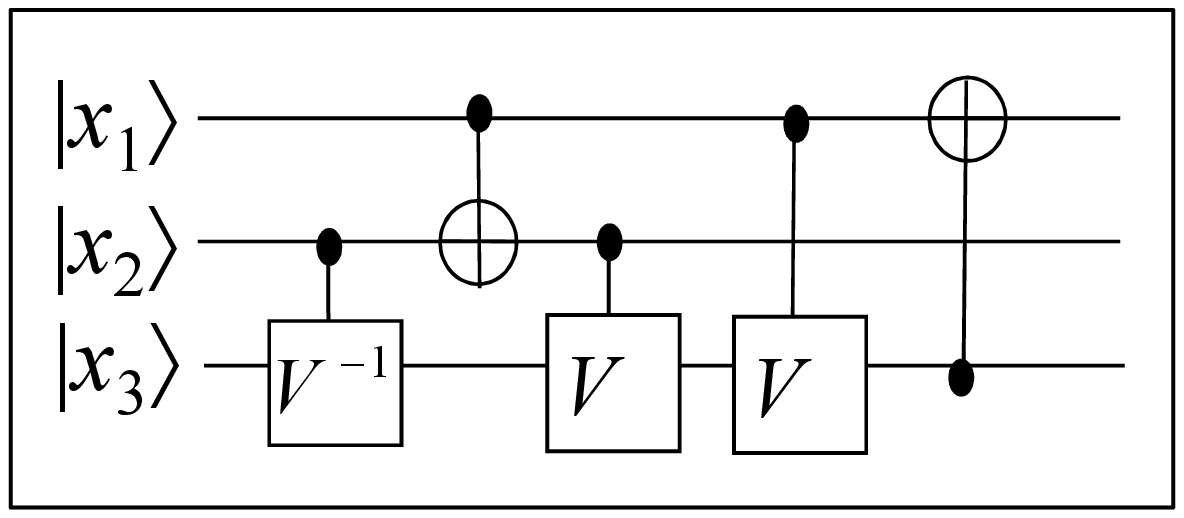}
}
\caption{A quantum circuit for half-adder (SCQC)}\label{fig:adder}
\end{center}
\end{minipage}
\begin{minipage}{5cm}
\begin{center}
 \resizebox{!}{1.5cm}{
 \includegraphics*{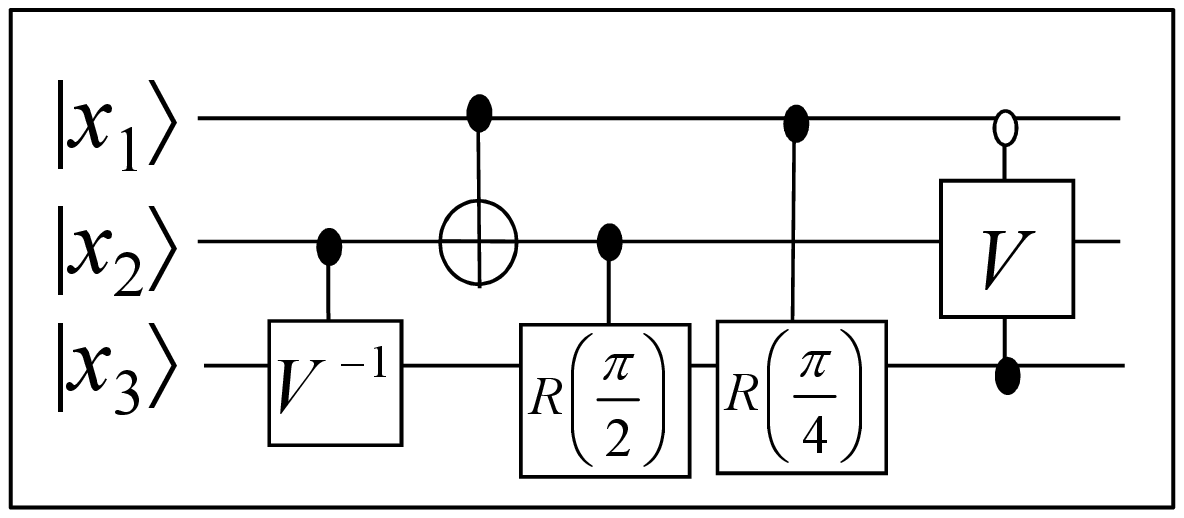}
 }
 \caption{A non-SCQC}\label{fig:NON-SCQC}
\end{center}
\end{minipage}
\end{figure}

Consider Fig.~\ref{fig:Circuit1} again. This circuit transforms the state
of the third bit  $\left|x_3\right>$  into  $\left|x_3 \oplus f(x_1,
  x_2)\right>$, where $f(x_1, x_2) = x_1 \cdot x_2 + \overline{x_1}
\cdot \overline{x_2}$. 
(Throughout the paper, we use $\overline{F}$ to mean the logical
negation of $F$.) 
Thus, we can use this circuit (as a part of a quantum algorithm) to
calculate the Boolean function $f(x_1, x_2) = x_1 \cdot x_2 + 
\overline{x_1} \cdot \overline{x_2}$. 
As mentioned before, although our goal is to construct such a quantum
circuit that calculates a Boolean function, quantum specific gates
(such as $R(\theta)$ and $V$) are useful~\cite{AP06,MYM05,HSYYP04,cuccaro-2004} to
make the circuit size smaller. 
 For example, the circuit as shown in Fig.~\ref{fig:adder} (reported
 in~\cite{HSYYP04}) utilizes controlled-$V$ and controlled-$V^{-1}$ 
 gates to become much smaller than the best one with only
 classical type gates, i.e., controlled-$NOT$ gates. (That was confirmed by 
 an essentially exhaustive search~\cite{HSYYP04}.)


In order to characterize such a quantum circuit that calculates a classical Boolean function 
with non-classical gates, we introduce a {\it Semi-Classical Quantum
  Circuit (SCQC)} whose definition is as follows.  

\begin{definition}
A Semi-Classical Quantum Circuit (SCQC) is a quantum circuit
consisting of controlled-U gates with the following restriction. 

{\bf Restriction.} 
If all the initial input quantum states of the circuit are
$\left|1\right>$ or $\left|0\right>$ (i.e., just classical values),
the quantum states of the control qubits of all the gates in the circuit
should be $\left|1\right>$ or $\left|0\right>$ at the time when the
gate is being operated. 
\end{definition}

The circuit in Fig.~\ref{fig:adder} is an SCQC. 
This is because the quantum states of the control qubits of all the gates are either
 $\left|1\right>$ or $\left|0\right>$ when the gate is being operated if  
 the initial input states $\left|x_1\right>$, $\left|x_2\right>$ and
 $\left|x_3\right>$ are either $\left|1\right>$ or $\left|0\right>$. 
It is not trivial to see the condition for  the quantum state of the
control qubit of the last gate (i.e., $\left|x_3\right>$) in
Fig.~\ref{fig:adder}. However, by using our new concepts (explained in
the next section), it is easy to verify that the state is indeed the
classical value if the input states of the circuit are classical values.

 On the contrary, the circuit as shown in Fig.~\ref{fig:NON-SCQC} is not an
 SCQC. Again, by using our new concept it is easily verified that the
 condition is not satisfied for the quantum state of the control qubit of
 the last gate (i.e., $\left|x_3\right>$) in Fig.~\ref{fig:NON-SCQC}. 


Our motivation to introduce SCQCs is based on the following observations. 

\begin{itemize}
\item Although SCQCs are in a subset of all the possible quantum
circuits, quantum circuits (for calculating a Boolean function)
designed by the  existing methods are all SCQCs to the best of our
knowledge
~\cite{HSYYP04}.   

\item Even in the future, it is very unlikely that we come up with 
  a {\it tricky} design method that produces a non-SCQC to  
  calculate (classical) Boolean functions. The reason is as follows. 
 If the circuit is not an SCQC, there is a gate such that the quantum
 state of its control bit is not a simple classical value 
 ($\left|0\right>$ nor $\left|1\right>$). In such a case, the quantum
  states of the control bit and the target bit after the gate cannot be
  considered separately: their states are not only non-classical 
  values but also correlated with each other. Such a situation is
  called quantum {\it superposition} and {\it
    entanglement}~\cite{Nielsen2000}. Since the 
  whole circuit should calculate a classical Boolean function, all of 
  the final output quantum states should be again restored to simple 
  classical values (i.e., $\left|0\right>$ or $\left|1\right>$) 
 if all the initial input quantum states of the circuit are simple
 classical values.  
  The reverse operations of creating quantum superposition and
  entanglement seems to be the only method to restore to a simple
  classical value. Thus, it seems nonsense to consider non-SCQC
  circuits when we consider practical design methods of quantum
  circuits to calculate Boolean functions. 
\end{itemize}

{\bf Important Note:} 
The restriction of SCQCs means that we cannot make {\it  
 entanglement} if all the initial input quantum states of an SCQC 
 are just classical values. It is well-known that quantum computation
 without entanglement has no advantage over classical computation. 
However, this does not mean that SCQCs are meaningless by the 
following reason: As mentioned, an SCQC is used as a sub-circuit to
calculate a Boolean function for some quantum algorithms. Thus, in the
real situation where an SCQC is used as a sub-circuit, the inputs to
the SCQC are not simple classical values, and so it indeed creates 
entanglement which should give us the advantage of quantum
computation. 
In other words, the restriction of SCQCs in the definition is 
considered when we suppose the inputs of SCQCs are just classical
values, which is not a real situation where SCQCs are really used. 


Therefore, SCQCs should be enough if we consider designing a quantum circuit to
calculate a Boolean function from the practical point of view. 
Moreover, the restriction of SCQCs 
provides us  an efficient method to analyze and verify quantum circuits 
as we will see in Sec.~\ref{sec:veri}. That is our motivation to
introduce the new concept in this paper.

\subsection{Quantum Functions and Matrix Functions}

Before introducing our new representation of the functionalities of
SCQCs, we need the following definitions.

\begin{definition}
A quantum function with respect to $n$
Boolean variables $x_1, x_2, \cdots, x_n$ is a mapping from $\{0,
1\}^n$ to qubit states.
\end{definition}

See the third bit after the first gate in the circuit in
Fig.~\ref{fig:adder} again. 
If the initial state of $\left|x_3\right>$ is $\left|0\right>$, the 
resultant state of the third bit can be seen as a quantum function 
described as $qf_1(x_1, x_2)$  in the second column of Table~\ref{TB}.
For example, the resultant quantum state becomes
$V^{-1} \left|0\right>$  when $x_1 = 0, x_2 = 1$. Thus,
$qf_1(0, 1)$ is defined as $V^{-1} \left|0\right>$ as shown in the table.

\begin{table}[tb]
\begin{center}
\caption{A truth tables for quantum, classical and matrix functions}\label{TB}
\scalebox{0.8}{
 \begin{tabular}{|c|c|c|c|c|c|c|} \hline
$x_1, x_2$ & $qf_1$  & $qf_2$  & $mf_1$ & $mf_2$ &  $CM(I)$  &  $CM(R(\frac{1}{2}\pi))$  \\ \hline
$0, 0$ & $\left|0\right>$ & $\left|1\right>$ &     $I$  &         $NOT$  & $I$       &   $R(\frac{1}{2}\pi)$      \\ \hline
$0, 1$ & $V^{-1} \left|0\right>$  & $\left|0\right>$  & $V^{-1}$&  $I$ & $I$       &   $R(\frac{1}{2}\pi)$      \\ \hline
$1, 0$ & $\left|0\right>$  & $\left|0\right>$  & $I$  &            $I$ &  $I$       &   $R(\frac{1}{2}\pi)$      \\ \hline
$1, 1$ & $V^{-1} \left|0\right>$  & $\left|1\right>$ & $V^{-1}$&   $NOT$ &  $I$       &   $R(\frac{1}{2}\pi)$      \\ \hline
\end{tabular}
}
\end{center}
\end{table}

Note that a Boolean function can be seen as a special case of quantum
functions. For example, the third column ($qf_2$) of Table~\ref{TB}
shows the quantum function of the resultant third qubit 
after the two gates of the circuit in Fig.~\ref{fig:Circuit1} when the initial state of
$\left|x_3\right>$ is $\left|0\right>$. This can be considered as the
output of a Boolean function when $\left|0\right>$  and 
$\left|1\right>$  are considered as Boolean values $0$ and $1$,
respectively. (As mentioned before, the circuit is considered to
calculate the Boolean function: $x_1 \cdot x_2 + \overline{x_1} \cdot 
\overline{x_2}$, which we consider essentially the same as 
($qf_2$) in Table~\ref{TB}.) 

The value of a quantum function $q(x_1, x_2, \cdots, x_n)$ can
always be expressed as $mf(x_1, x_2, \cdots, x_n) \left|0\right>$,
where $mf(x_1, x_2, \cdots, x_n)$  is a mapping from $\{0, 1\}^n$
to  2$\times$2 unitary matrices. It is convenient to consider $mf(x_1,
x_2, \cdots, x_n)$ instead of $q(x_1, x_2, \cdots, x_n)$ itself,
thus we introduce the following definition.

\begin{definition}
A matrix function with respect to $n$ 
Boolean variables $x_1, x_2, \cdots, x_n$ is a mapping from $\{0,
1\}^n$ to  2$\times$2 (unitary) matrices.
\end{definition}

The fourth and the fifth columns of Table~\ref{TB} 
show  the matrix function $mf_1$ and $mf_2$ for the quantum 
function $qf_1$ and $qf_2$, respectively, in the same table.
In this paper, we treat a matrix function whose output values are
only $I$ or $NOT$ as a classical Boolean function by considering that
$NOT$ and $I$ of the matrix function correspond to $1$ and $0$, 
respectively, of the Boolean function.
In other words, we represent a Boolean function by a matrix function
as a special case. 

We define a special type of matrix function called {\it constant
  matrix function} as follows.

\begin{definition}
A matrix function $mf(x_1, x_2, \cdots, x_n)$
is called a constant matrix function if $mf(x_1, x_2, \cdots, x_n)$
 are the same for all the assignments to $x_1, x_2, \cdots, 
 x_n$. $CM(M)$ denotes a constant matrix function that
always equals to the matrix $M$.
\end{definition}

The sixth and the seventh columns of Table~\ref{TB} show the truth
tables for constant matrix functions, $CM(I)$ and
$CM(R(\frac{1}{2}\pi))$, respectively.

\begin{table}[tb]
\begin{center}
\caption{Operators $\oplus$ and $\ast$.}    \label{operators}
\scalebox{0.9}{
 \begin{tabular}{|c|c|c|c|c|c|c|} \hline
$x_1, x_2$ & $mf_1$     &   $mf_2$  & $mf_1 \oplus mf_2$   & $mf_3$     &   $f$       & $f \ast mf_3$      \\ \hline  
$0, 0$     & $R(\frac{1}{2}\pi)$ &  $R(\frac{1}{2}\pi)$    & $R(\pi)$   & $R(\frac{1}{2}\pi)$ &  1   & $R(\frac{1}{2}\pi)$      \\ \hline          
$0, 1$     & $I$        &  $I$                             & I           & $I$                 &  0   & I               \\ \hline
$1, 0$     & $I$        &   $R(\frac{1}{4}\pi)$            & $R(\frac{1}{4}\pi)$  & $R(\pi)$           &  1   & $R(\pi)$               \\ \hline
$1, 1$     & $R(\frac{1}{2}\pi)$ & $R(\frac{1}{4}\pi)$     & $R(\frac{3}{4}\pi)$  & $R(\pi)$            &  0   & I              \\ \hline
\end{tabular}
}
\end{center}
\end{table}


By using the matrix function $mf_1$ in the fourth column of
Table~\ref{TB}, we can easily see how the first gate  in
Fig.~\ref{fig:adder} transforms 
the third qubit $\left|w\right>$: $\left|w\right>$ is transformed
to $mf_1(x_1, x_2) \left|w\right>$. For example, when $x_1 = 0,
x_2 = 1$, $\left|w\right>$ is transformed to $V^{-1} \left|w\right>$.

We would like to stress again the following point:  The above means
that the representation (and so the analysis) by  
matrix functions works even when $\left|w\right>$ is any general
quantum state. Indeed,  we can  use an SCQC even when the input states
are not simple classical values, i.e., the restriction of SCQCs does
not say that SCQCs cannot be used when the inputs are not
classical. (If so, we may not be able to use an SCQC for a part of
a quantum algorithm.) 

For matrix functions, 
we introduce two operators ``$\oplus$'' and `$\ast$,'' which are used
to construct DDMFs for a quantum circuit in the following sections. 

\begin{definition}\label{def1}
Let $mf_1$, $mf_2$ and $mf_3$ be matrix functions with respect to $x_1$ to
$x_n$. Then $mf_1 \oplus mf_2$ is defined as a matrix
function $mf$ such that $mf(x_1, \cdots, x_n) = mf_1(x_1, \cdots, x_n)
\cdot mf_2(x_1, \cdots, x_n)$ where $\cdot$ means normal matrix
multiplication. Let also $f$ be a  Boolean function with respect to
$x_1$ to $x_n$. Then 
 $f \ast mf_3$ is a matrix function which equals to $mf_3(x_1, x_2,
 \cdots, x_n)$ when  $f(x_1, x_2, \cdots, x_n)=1$, and equals to $I$ 
 when $f(x_1, x_2, \cdots, x_n)=0$.
\end{definition}

Note that the operator $\ast$ is defined as asymmetric, i.e., the first
argument should be a Boolean function whereas the second argument can
be any matrix function. 
This is due to the restriction of SCQCs such that the state of a 
control bit should be $\left|1\right>$ or $\left|0\right>$ (i.e.,
just classical value) whereas the state of a target bit can be any
quantum state.

See examples in Table~\ref{operators}. Note that if both of $mf_1$ and 
$mf_2$ are considered to be Boolean functions like $mf_2$ in  
Table~\ref{TB},  the operator $\oplus$ corresponds to the EXOR of the
two  Boolean functions.  
Note also that if $mf_3$ is essentially a Boolean function like 
$mf_2$ in Table~\ref{TB}, the operator $\ast$ corresponds to the AND 
of the two Boolean functions.

\subsection{Decision Diagrams for Matrix Functions}

\begin{figure}[tb]

\begin{minipage}{8cm}
\begin{center}
\resizebox{2cm}{!}{
\includegraphics*{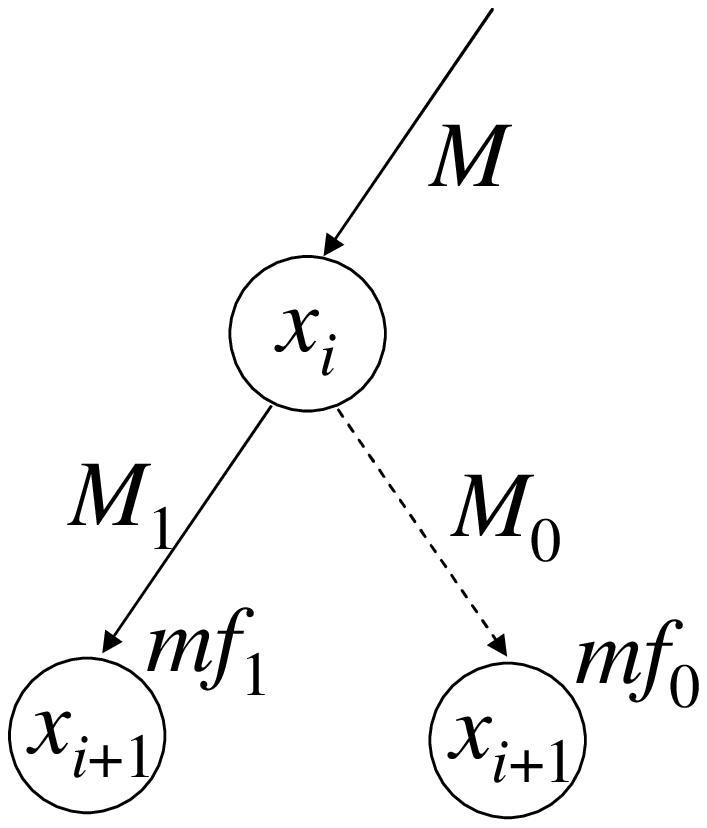}
}
\caption{An internal DDMF node}\label{fig:DDMF1}
\end{center}
\end{minipage}
\begin{minipage}{8cm}
\begin{center}
\resizebox{4cm}{!}{
\includegraphics*{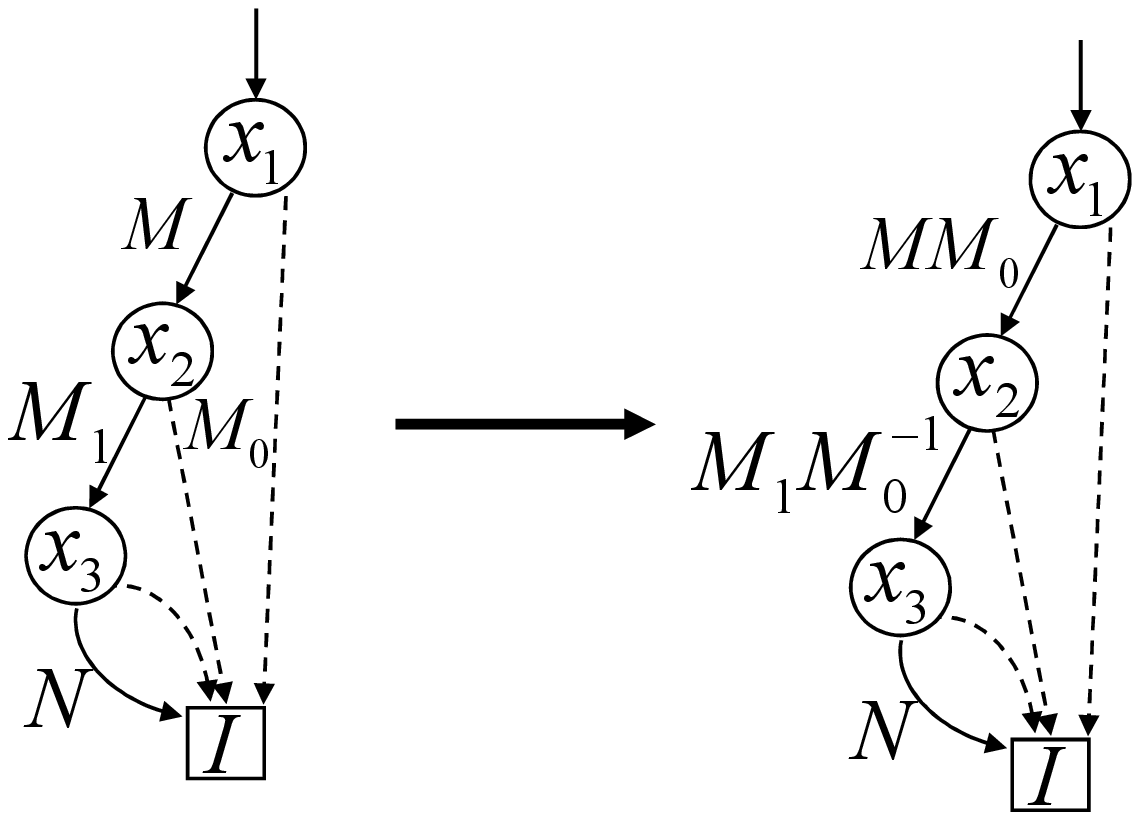}
}
\caption{Conversion to the canonical form}\label{fig:canonicali}
\end{center}
\end{minipage}
\end{figure}

A matrix function for a quantum function can be
expressed efficiently by using an edge-valued binary decision diagram
structure, which we call a DDMF whose definition is as follows: 

\begin{definition}
A  Decision Diagram for a Matrix Function (DDMF) is a directed acyclic
graph with three types of nodes: (1) A single terminal node
corresponding to the identity matrix $I$, (2) a root node with an
incoming edge having a weighted matrix $M$, and (3) a set of
non-terminal (internal) nodes. 

Each internal and the root node are associated with a Boolean 
 variable $x_i$, and have two outgoing edges which are called 1-edge
 (solid line) leading to another node (the 1-child node) and 0-edge
 (dashed line) leading to another node (the 0-child node).
 Every edge has an associated matrix.

The matrix function represented by a node is defined recursively by
the following three rules.

(1) The matrix function represented by the terminal node is the
  constant matrix function $CM(I)$.

(2) The matrix function represented by an internal node (or the root
   node) whose associated variable is $x_i$ is defined as
   follows:  $x_i \ast (CM(M_1) \oplus mf_1) \oplus \overline{x_i}
   \ast (CM(M_0) \oplus mf_0)$, where $mf_1$ and $mf_0$ are the matrix functions represented by
   the 1-child node and the 0-child node, respectively, and
  $M_1$ and $M_0$ are the matrices of the 1-edge and the 0-edge,
   respectively. (See an illustration of this structure in
   Fig.~\ref{fig:DDMF1}.) 

(3) The root node has one incoming edge that has a matrix
   $M$. Then the matrix function represented by the whole DDMF is
   $CM(M) \oplus mf$, where $mf$ is a matrix function represented by the
   root node.

\end{definition}

Like conventional BDDs, we achieve the canonical form for a DDMF if we
impose the following restriction on the matrices on all the edges.

\begin{definition}
A (DDMF) is canonical when 
 (1) all the matrices on 0-edges are $I$,
 (2) there are no redundant nodes: No node has 0-edge and
   1-edge pointing to the same node with $I$ as the 1-edge matrix, 
 and (3) common sub-graphs are shared: There are no two identical sub-graphs. 
\end{definition}

Any DDMF can be converted to its canonical form by using the following
transformation from the terminal node to the root node: Suppose the
matrices on incoming edge, 0-edge and 1-edge of a node be 
$M$, $M_0$  and $M_1$, respectively. Then, if $M_0$ is not $I$, we
modify these three matrixes as follows: 
(1) The matrix on the incoming edge is changed to be  $M M_0$. 
 (2) The matrix on the 1-edge is changed to be  $M_1 M_0^{-1}$.
 (3) The matrix on the 0-edge is changed to be  $I$. 
It is easily verified that this transformation does not change the
matrix function represented by the DDMF. See the example in
Fig.~\ref{fig:canonicali} where the matrix on 0-edge of the node
$x_2$ is converted to $I$. In the example, the matrices $I$ on
edges are omitted.

{\bf Note:} 
The concepts of {\it quantum functions} and {\it matrix functions}
 may be used implicitly in the design method of~\cite{AP06}, and the
 decision diagram structure is similar between DDMFs and the quantum
 decision diagrams used in~\cite{AP06}.  
However, the quantum decision diagrams in~\cite{AP06} are used to represent 
conventional Boolean functions whereas  DDMFs are used for 
representing matrix functions: the terminal node of a DDMF is a 
matrix $I$. Also a weight on an edge in DDMFs is generalized to 
any matrix. Thus,  DDMFs can be considered as a generalization of 
quantum decision diagrams to treat matrix functions rather than 
Boolean functions. (As we have seen in Table~\ref{TB}, Boolean
functions can be seen as a special case of quantum functions.)


We will use the same operators, $\oplus$ and $\ast$, for DDMFs as for
matrix functions: 
\begin{definition} 
Let $DDMF_{mf_1}$, $DDMF_{mf_2}$ and $DDMF_{mf_3}$ be DDMFs that 
represent matrix functions $mf_1$, $mf_2$ and $mf_3$, respectively. 
Then $DDMF_{mf_1} \oplus DDMF_{mf_2}$ 
 is defined as a DDMF that represents a matrix function $mf_1 \oplus
 mf_2$.  
Let also $DDMF_{f}$ be a DDMF that represents  a Boolean function $f$. 
Then $DDMF_{f} \ast DDMF_{mf_3}$  
 is defined as a DDMF that represents a matrix function $f \ast mf_3$
\end{definition}

\section{Verification of SCQCs by using DDMFs}\label{sec:veri}

\begin{figure}[tb]
\begin{minipage}{8cm}
\begin{center}
\resizebox{3cm}{!}{
\includegraphics*{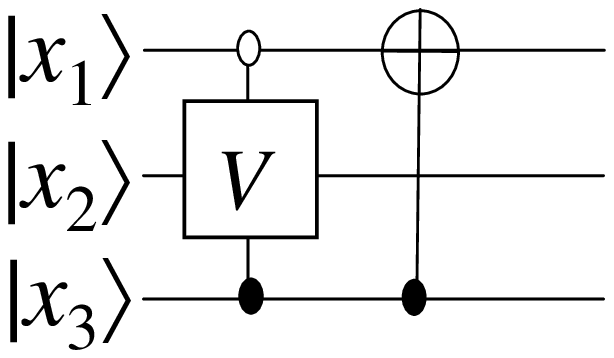}
}
\caption{An SCQC (1)}\label{fig:SCQC1}
\end{center}
\end{minipage}
\begin{minipage}{8cm}
\begin{center}
\resizebox{3cm}{!}{
\includegraphics*{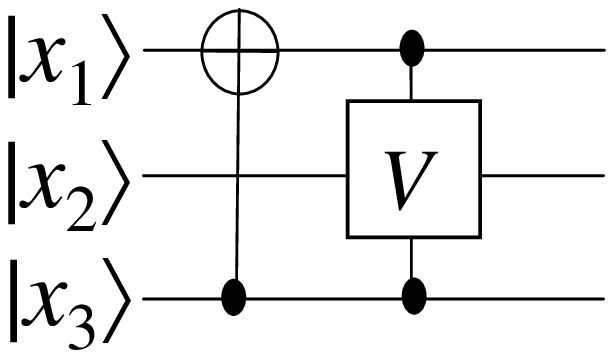}
}
\caption{An SCQC (2)}\label{fig:SCQC2}
\end{center}
\end{minipage}
\end{figure}

See two SCQCs in Fig.~\ref{fig:SCQC1} and Fig.~\ref{fig:SCQC2}. 
It is easy to see that their functionalities are the same. However, the
problem is how to verify the equality for much larger circuits. 
Thanks to the introduction of DDMFs, we propose a method to verify the
equality of given two $n$-qubit SCQCs in the following. 



{\noindent \bf Step 1.} We construct a DDMF to represent the matrix function that
   expresses the functionality for each qubit state at the end of each circuit. 

{\noindent \bf Step 2.} We compare two DDMFs for the corresponding
qubits of the 
   two circuits. The comparison of two DDMFs can be done in $O(1)$ time
   as in the case of BDDs. 

Step.~1 is performed in a similar manner of constructing BDDs to 
represent each Boolean function in a logic circuit: (1) We first
construct  a DDMF for each primary input $x_i$, and then (2) we pick a
gate one by one from the primary inputs, and construct a DDMF for the
output function of the gate from DDMFs for the input functions of the
gate. The construction of a DDMF from two DDMFs can be done
recursively as exactly the same as the construction of a BDD from two  
BDDs~\cite{brya86}.  
In the below, we use a notation $D_{i}^{j}$ to express the DDMF for
the $i$-th 
quantum qubit state right after the $j$-th gate. 
We also use a notation $F(D)$ to denote the matrix function (or the
Boolean function in a special case) represented by a DDMF $D$. 

{\noindent \bf Initialization.} 
 For each input $x_i$, we construct a $D_{i}^{0}$ as a
  DDMF for $x_i$. This is the DDMF for the matrix function
  (in fact, essentially a Boolean function) which is $NOT$ when $x_i=1$.

{\noindent \bf Construction of the DDMFs right after the $j$-th gate.} 
 From the first gate to the last gate,  
  we construct $D_{i}^{j}$ from $D_{i}^{j-1}$ as follows. 
  If the $i$-th bit is not the target bit of the $j$-th gate, $D_{i}^{j} =
  D_{i}^{j-1}$. 
  If the $i$-th bit is the target bit of the $j$-th gate
  $D_{i}^{j} =  D_{i}^{j-1} \oplus D_{gate}$ where
  $D_{gate}$ is constructed by the following two steps. 

(1) For the $j$-th gate, let us suppose that the positive control bits be the
$p_1, p_2, \cdots,  p_k$-th bits, and the negative control bits be the $n_1, n_2,
 \cdots, n_l$-th bits. 
 Then, by the restriction of SCQCs, all the matrix functions 
   $F(D_{m}^{j-1})$ for $m = p_1, p_2, \cdots,
    p_k, n_1, n_2, \cdots, n_l$ are essentially classical Boolean functions. 
(Therefore, in the following expression, we treat $F(D_{m}^{j-1})$ as
 Boolean functions, and perform logical operations on them.) 
Thus we can calculate a logical AND of them: 
{\small $g = F(D_{p_1}^{j-1}) \cdot F(D_{p_2}^{j-1}) \cdots F(D_{p_k}^{j-1})
\cdot \overline{F(D_{n_1}^{j-1})} \cdot  \overline{F(D_{n_2}^{j-1})}
 \cdots \overline{F(D_{n_l}^{j-1})}$.}
Note that this Boolean function can be obtained by DDMF operations
since a DDMF represents a Boolean function in a special case. 

(2) We construct $D_{gate} = (DDMF$ for $g$) $\ast (DDMF$ for $CM(U))$, 
    where $U$ is a unitary matrix associated with the $j$-th gate.

Note that all the DDMF operations in the above should be performed
 efficiently by using {\it Apply} operations and {\it operation and
   node hash tables} as the  conventional BDD operations~\cite{brya86}.

\begin{figure}[tb]
\begin{minipage}{8cm}
\begin{center}
\resizebox{4cm}{!}{
\includegraphics*{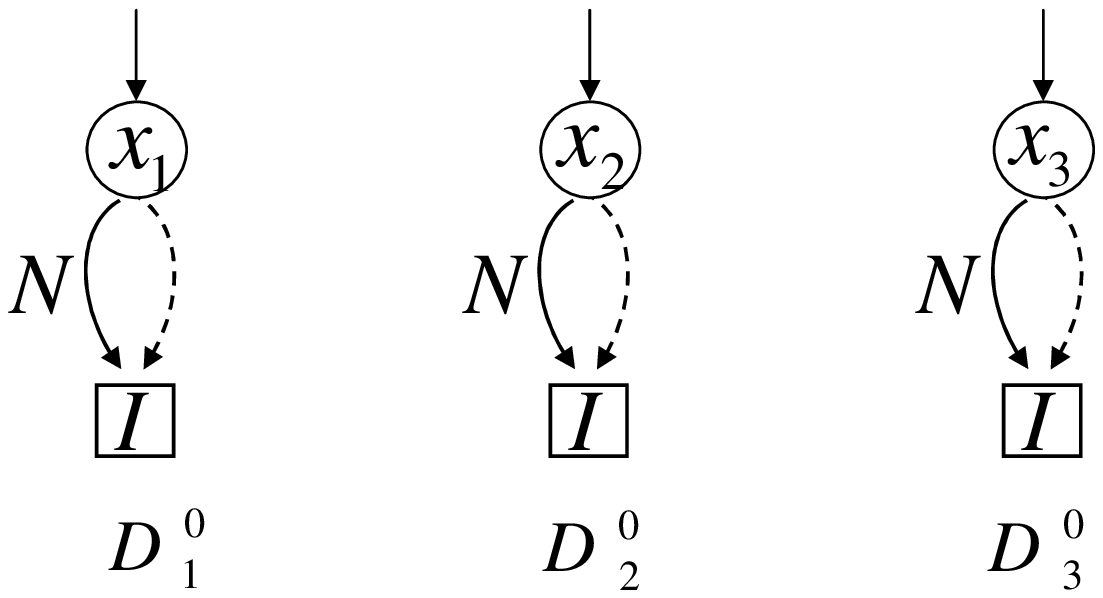}
}
\caption{DDMFs after the initialization}\label{fig:spte0}
\end{center}
\end{minipage}
\hspace*{0.3cm}
\begin{minipage}{8cm}
\begin{center}
\resizebox{3.8cm}{!}{
\includegraphics*{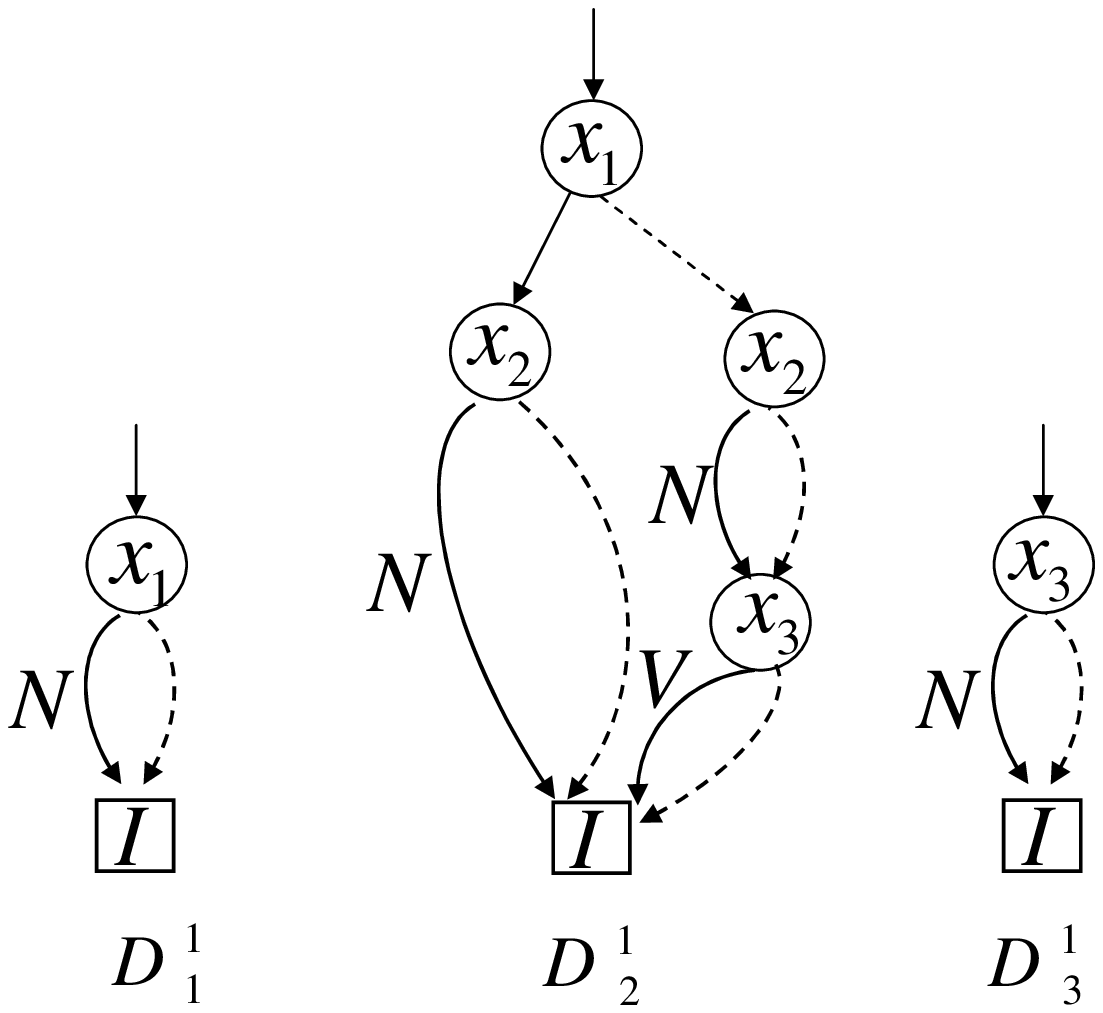}
}
\caption{DDMFs after a gate}\label{fig:spte1}
\end{center}
\end{minipage}
\end{figure}

We show an example of DDMFs for the quantum circuit as shown in
Fig.~\ref{fig:SCQC1}.  
At the initialization step, we construct DDMFs for functions, $x_1,
x_2$ and $x_3$, which are $D_{1}^{0}$, $D_{2}^{0}$ and
$D_{3}^{0}$, respectively,  as shown in Fig.~\ref{fig:spte0}. 
Then we construct the DDMFs for the quantum states 
right after the first gate. Since the target bit is the second bit for
the first gate, $D_{1}^{1} =  D_{1}^{0}$, and  $D_{3}^{1} =
D_{3}^{0}$. To construct $D_{2}^{1}$, we first calculate 
a Boolean function $g = \overline{F(D_{1}^{0})} \cdot F(D_{3}^{0})
= \overline{x_1} \cdot x_3$. 
This is because the first bit and the third bit are negative and
 positive controls, respectively. 
Then we construct $D_{gate} = (DDMF$ for $g$) $\ast (DDMF$ for $CM(V))$, whose
matrix function is shown in Table~\ref{DDMF-control}. 
Finally, we construct 
  $D_{2}^{1} =  D_{2}^{0} \oplus D_{gate}$ whose 
matrix function is as shown in Table~\ref{DDMF-2}. 
The constructed DDMFs after the first gate are 
shown in Fig.~\ref{fig:spte1}.

\begin{table}[tb]
\begin{minipage}{8cm}
  \begin{center}
\caption{\hspace*{-2mm}A truth table for $F(D_{gate})$}
    \label{DDMF-control}
 \begin{tabular}{|c|c|} \hline
$x_1, x_2, x_3$ & $F(D_{gate})$  \\ \hline
$0, 0, 0$ & $I$  \\ \hline
$0, 0, 1$ & $V$  \\ \hline
$0, 1, 0$ & $I$  \\ \hline
$0, 1, 1$ & $V$  \\ \hline
$1, 0, 0$ & $I$  \\ \hline
$1, 0, 1$ & $I$  \\ \hline
$1, 1, 0$ & $I$  \\ \hline
$1, 1, 1$ & $I$  \\ \hline
\end{tabular}
\end{center}
\end{minipage}
\begin{minipage}{8cm}
  \begin{center}
\caption{A truth table for $F(D_{2}^{1})$}
    \label{DDMF-2}
 \begin{tabular}{|c|c|} \hline
$x_1, x_2, x_3$ & $F(D_{2}^{1})$  \\ \hline
$0, 0, 0$ & $I$  \\ \hline
$0, 0, 1$ & $V$  \\ \hline
$0, 1, 0$ & $N$  \\ \hline
$0, 1, 1$ & $VN$  \\ \hline
$1, 0, 0$ & $I$  \\ \hline
$1, 0, 1$ & $I$  \\ \hline
$1, 1, 0$ & $N$  \\ \hline
$1, 1, 1$ & $N$  \\ \hline
\end{tabular}
\end{center}
\end{minipage}
\end{table}

\section{Comparison with the previous methods} 

In this section we compare our method with the previous
methods to show the advantage of our method.

\subsection{Verification method based on previous techniques} 

\begin{figure*}[tb]
\begin{minipage}{8cm}
\begin{center}
\resizebox{5cm}{!}{
\includegraphics*{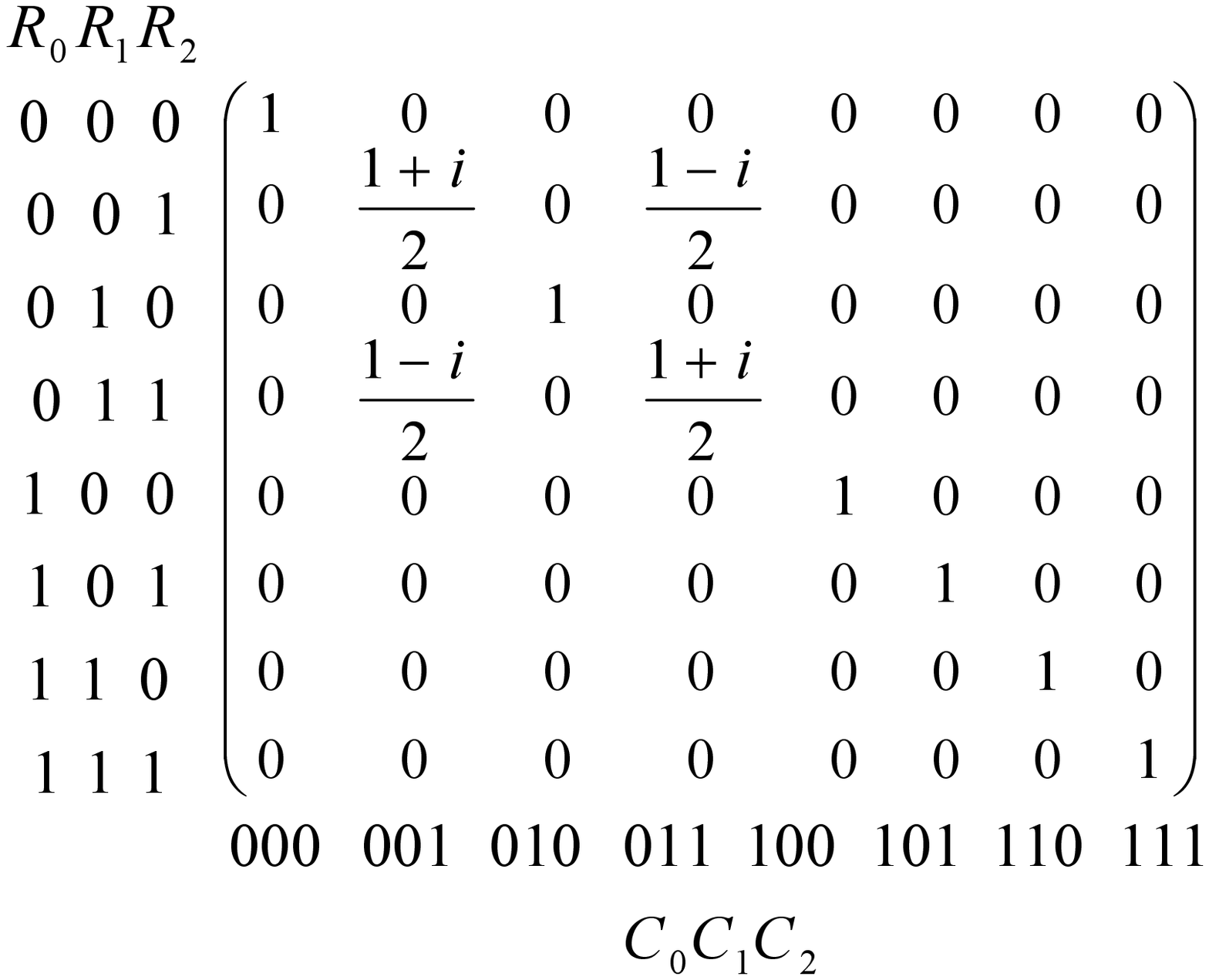}
}
\caption{A Unitary Matrix for a 3-qubit gate}\label{fig:matrix2}
\end{center}
\end{minipage}
\begin{minipage}{5cm}
\begin{center}
\resizebox{4.05cm}{!}{
\includegraphics*{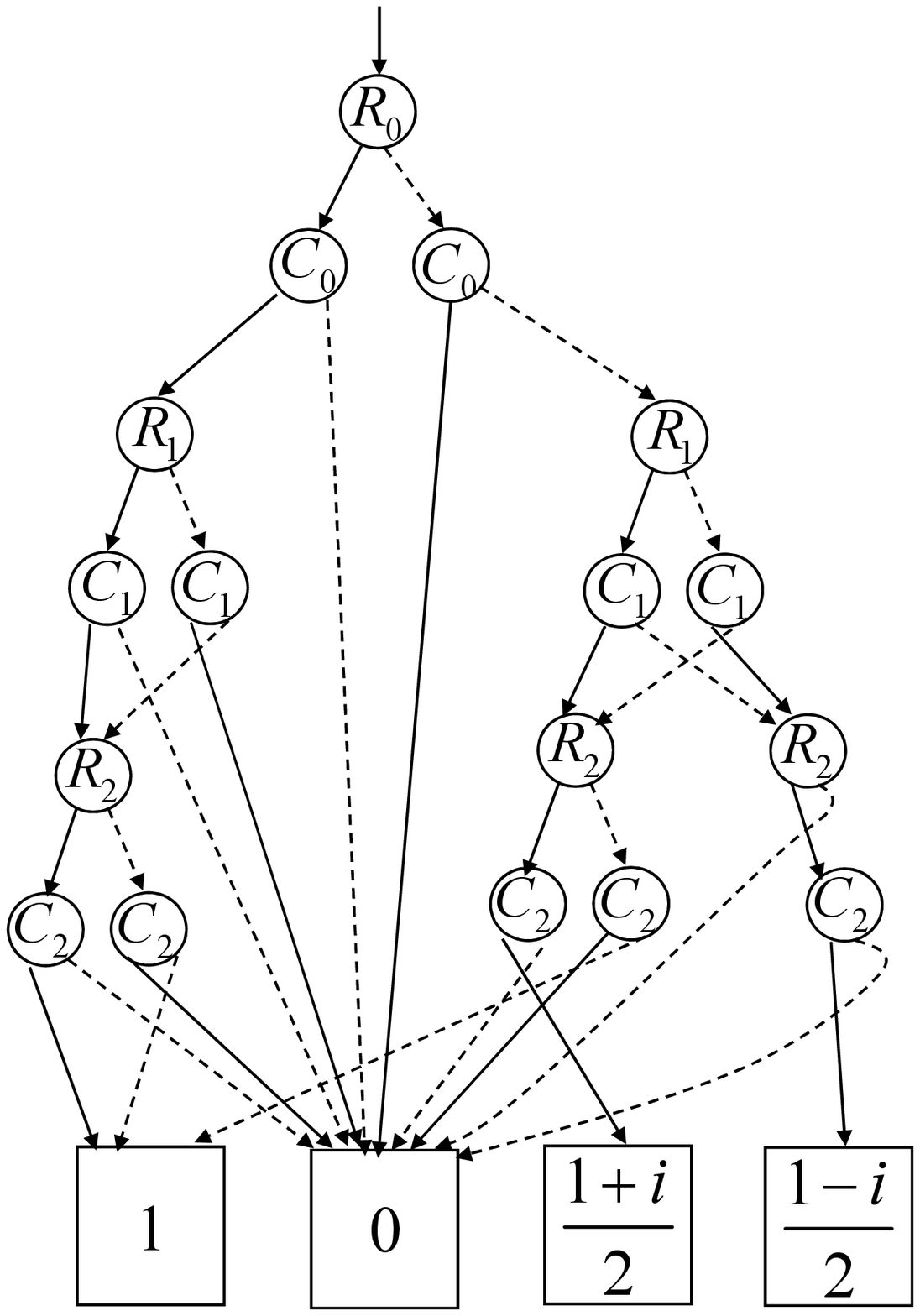}
}
\caption{A QuIDD (1)}\label{fig:QuIDD1}
\end{center}
\end{minipage}
\begin{minipage}{7cm}
\begin{center}
\resizebox{1.5cm}{!}{
\includegraphics*{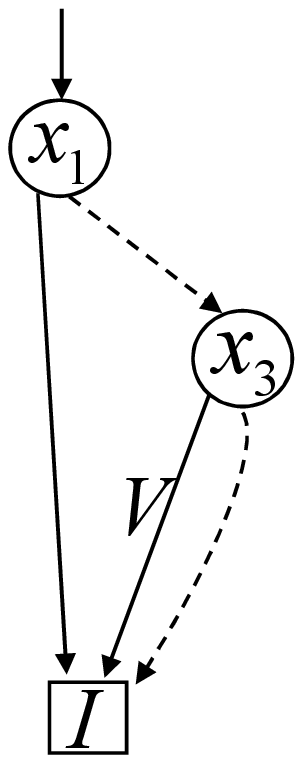}
}
\caption{A DDMF}\label{fig:DDMF2}
\end{center}
\end{minipage}
\begin{minipage}{6cm}
\begin{center}
\resizebox{5cm}{!}{
\includegraphics*{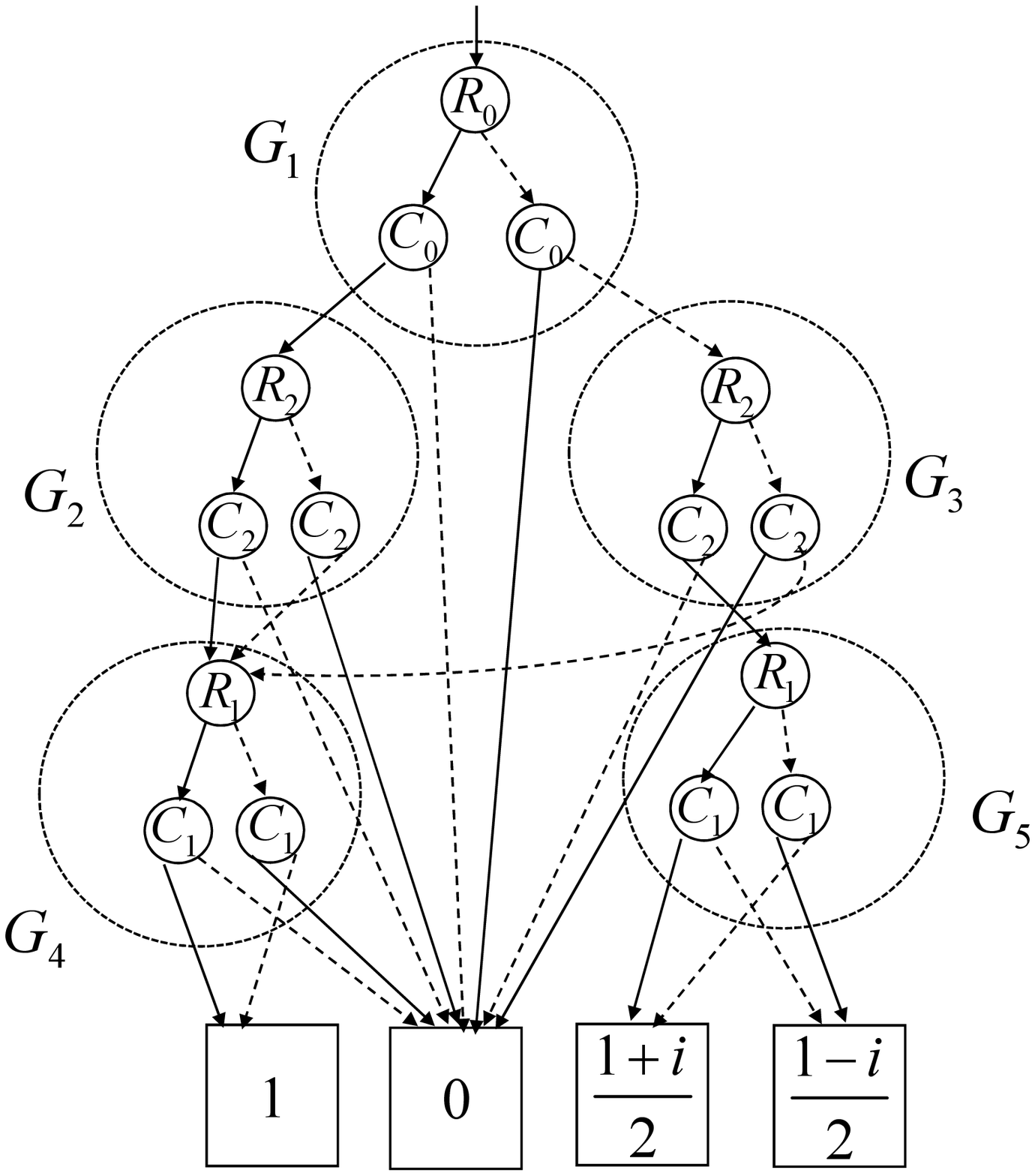}
}
\caption{A QuIDD (2)}\label{fig:QuIDD2}
\end{center}
\end{minipage}

\end{figure*}

A gate (or a circuit) of $n$ qubits can be described by a 
 $2^n \times 2^n$ unitary matrix. 
 For example,  the unitary matrix that expresses the functionality of the
 last gate in Fig.~\ref{fig:NON-SCQC} can be shown as in
 Fig.~\ref{fig:matrix2}. 

Since the same structure (sub-matrices) are often repeated 
 in such a $2^n \times 2^n$ unitary matrices (like 
 Fig.~\ref{fig:matrix2}),  data compression
 schemes based on decision  diagram structures have been proposed: 
QuIDDs~\cite{VMH05} based on multi-terminal binary decision diagrams, and 
QMDDs~\cite{MT06} based on multiple-valued decision diagrams 
(MTBDDs)~\cite{CFZ96}. 
Here we explain how the data compression works for QuIDDs.  
QMDDs have a slightly different approach: They use multi-valued logic
instead of binary logic, and the strategy of selecting decision 
variables is a bit different. However, it should be noted that the two  
approaches are essentially the same when (1) the target circuit is
binary logic ($\left|0\right>$ or $\left|1\right>$) valued (which is
our case), and (2) the variable ordering is appropriately chosen for
QuIDDs (as explained below).

A QuIDD for the matrix in Fig.~\ref{fig:matrix2} can be constructed as
shown in Fig.~\ref{fig:QuIDD1}. In a QuIDD representing a matrix, we 
have decision variables to specify rows ($R_0, R_1, R_2$) and columns
($C_0, C_1, C_2$) of the matrix as shown in
Fig.~\ref{fig:matrix2}. For example, the variable assignment 
to $R_0=0, R_1=1, R_2=1, C_0=0, C_1=0, C_2=1$) leads to the fourth
row and the second column element, $\frac{1-i}{2}$.  
We can construct a binary decision diagram where each variable 
assignment corresponds to one element of the matrix as shown in 
Fig.~\ref{fig:QuIDD1}. If there are some repeated structures in a
matrix, this diagram can reduce the necessary memory space to store
the matrix information. 

It is known that interleaving the row and the column variables 
(i.e., the order of $R_0, C_0, R_1, C_1, \cdots$) would be a good variable
order~\cite{VMH03}. In such a case, the variable order becomes the same
as in the case of QMDDs. In this paper, we also consider such a variable
order. 

As the conventional decision diagrams, we can implement any 
operations (such as addition and multiplication) between two QuIDDs
based on {\it Apply} operations and {\it operation and node hash
  tables}. Usually, QuIDDs can reduce the necessary memory, and the
necessary computational time for matrix operations for quantum circuit
simulations~\cite{VMH05}.

\subsection{Advantages of the Proposed Approach}

First we compare the number of nodes to represent the functionality of 
a single gate between DDMFs and QuIDDs. 
Let us use the last gate in
 Fig.~\ref{fig:NON-SCQC} for our explanation. As explained
 before, the QuIDD shown in Fig.~\ref{fig:QuIDD1} 
represents the matrix corresponding to the gate.  
A DDMF for the same purpose can be shown as in Fig.~\ref{fig:DDMF2}. 
In the DDMF, $V$ is attached on the path corresponding to the variable
assignment $x_1 =0, x_3 =1$. This is because the gate applies $V$ only 
when $x_1 =0, x_3 =1$.  

From the two figures, DDMFs seem to be better than 
QuIDDs. However, this is a bit unfair because the DDMF  
implicitly utilizes the fact that $x_2$ is the target bit. (Thus, $x_2$
does not appear in the DDMF.) On the other hand, the QuIDD does not
use such knowledge. 
Although the explanation is omitted (see the details in a standard
text such as~\cite{Nielsen2000}), $(R_1, C_1)$ corresponds to the  
input (and the output) line on $x_2$ in the original quantum 
circuit. Thus, if we know that $x_2$ is the target bit, we can also
choose the appropriate variable order for the QuIDD such that the 
pairs of variables $(R_1, C_1)$ are 
put on the bottom. Then, the QuIDD becomes smaller as shown in
Fig.~\ref{fig:QuIDD2}. 




Although there still seems to be a big difference between the two diagrams in
Figs.~\ref{fig:DDMF2} and ~\ref{fig:QuIDD2}, the essential 
difference is only a constant factor since we can decrease the
number of nodes of the QuIDD in Fig.~\ref{fig:QuIDD2} if we consider
the following two issues. 

(1) If we choose the appropriate variable ordering of $(R_i, C_i)$ as
described above, the corresponding matrix can be considered as one
such that there are only 2$\times$2 matrices on the diagonal. (Again we omit
the explanation.)  Thus, for each group of nodes, $(R_i, C_i)$, which
is not on the bottom (e.g., $G_1, G_2, G_3$ in
Fig.~\ref{fig:QuIDD2}), the two paths corresponding to $(R_i=1,C_i=0)$
and $(R_i=0,C_i=1)$ always go to $0$ 
terminal node. (This is because the matrix is a diagonal matrix,  and
thus the elements in the right upper and the left lower parts of the
matrix are all 0.) Thus, we can essentially omit such paths, and then only two
paths are essentially necessary for the group of nodes, which means
that we can replace each group of nodes by one node. (Of course, to do
so, we need  more operations than the standard QuIDD operations.) 

(2) If we note that the terminal nodes of a DDMF are 2$\times$2 matrices (which
has 4 elements), the last group of nodes ($G_4$ and $G_5$ in
Fig.~\ref{fig:QuIDD2}) should not be counted for the fair comparison.  

To sum up, in this example, $G_2$, $G_4$ and $G_5$ should not be
counted since each group leads to only the elements of a single 2$\times$2
matrix ($I$ or $V$), and each of $G_1$ and $G_3$ should be considered
as one 
node; thus there is no essential difference between the two diagrams. 
Of course, it is apparent that DDMFs are much more straightforward 
and easy to implement (hence should be faster) than QuIDD based 
approaches for our purpose. However, the above discussion makes it
clear that there is only a constant factor difference.  

Nevertheless, there is a good reason for us to introduce DDMFs:  
If we consider the verification of the two quantum circuits,   
the difference may become {\it exponential} in some cases, which will
be explained below. 



As mentioned in Sec.~\ref{sec:veri}, when we construct 
$D_{i}^{j}$ from the previous step, we always implicitly choose the 
appropriate variable order: We implicitly put the target bit (the
$i$-th bit) on the bottom (more precisely, we ignore the target bit)
when we calculate $D_{i}^{j}$. On the other hand,  we cannot choose an
appropriate variable ordering 
for the QuIDD approach since the verification by QuIDDs are performed
as follows: We can verify the equality of the two quantum   
circuits by comparing the two QuIDDs representing the two quantum 
circuits. To construct the QuIDD for a circuit, we simply multiply 
 matrices corresponding to gates in the circuit from the left to the 
right. This can be done by representing each matrix into a QuIDD. 
For the first gate, we can choose the appropriate variable order. 
However, if the target bits are different between the first and the  
second gates, we cannot choose the appropriate variable ordering when
we construct the QuIDD for the second gate. This is because 
the same variable order should be applied for the two QuIDDs when we 
 perform the multiplication. Thus, at least one of the QuIDDs may become
 much larger compared to the DDMF approach. 

Another important observation is that the resultant QuIDD after 
the multiplication may be larger than the corresponding DDMF approach
by the following reason. We construct each DDMF for each qubit, and
thus we can implicitly choose the best variable order (i.e., putting the
target bit to the bottom) for each qubit. 
This can be done because of the restriction of SCQCs. 
On the other hand, unitary matrix based approaches (such as QuIDD and
QMDD based verifications) do not assume such a restriction, and thus they
do not treat each qubit separately. Accordingly, they 
 represent the functionalities for all qubits at the same time as one
 unitary matrix corresponding to a part of quantum circuit. Therefore,
 we cannot choose a nice variable order if the appropriate variable order
 differs for different qubits. (This occurs when we multiply several
 matrices corresponding to quantum gates with different target bits.) 
Thus, in the worst case, QuIDDs become much larger than DDMFs during
the verification procedures. 
 It is obvious that the necessary memory and the necessary time for 
 {\it Apply} operations become smaller if the number of nodes becomes smaller. 
 Thus it is apparent that our approach is much more efficient  
 than previous approaches for the purpose of the verification of SCQCs.

It should be noted that there is also an apparent advantage of DDMFs
in terms of {\it operation and node hash tables} as follows. 
The variables for DDMFs during the verification is always the inputs
of the circuits (i.e., $x_1$ to $x_n$). In other words, we always represent
matrix functions with respect to {\it the inputs of the circuits.}
Thus we are always working on the {\it the input variables of the circuits}.
On the other hand, the variables for QuIDDs differs depending on the 
gates. More precisely, a unitary matrix corresponding to a gate (or a 
part of the circuit) represents a relation between the inputs of the 
 gate (or the part of the circuit)  and the outputs of that. That is,
 the variables for a QuIDD are always {\it local}: The meaning of
 variables $R_i$  and $C_i$ for a QuIDD changes during the matrix
 multiplication. (Even though we work on the same variables $R_i$ and
 $C_i$, the logical meaning of the variables changes depending on the
 corresponding quantum gates.) This apparently makes it difficult to   
 share the previously computed results in the hash tables. 
 Thus, it is apparent that hash tables work much better for DDMFs. 

The above discussion reveals why our verification method based on
DDMFs should work more efficiently than the previous approaches. 

We have implemented a DDMF library by C++, and performed a preliminary
experiment. Unfortunately, there is no large SCQC benchmark, and
thus we randomly generated SCQCs and constructed DDMFs for the generated
circuits. Then the average (of 10 trials) of the total number of
used nodes and the CPU time (on a Linux system running at 3.0 GHz with
256 MB memory) for various settings (i.e., the numbers of inputs and
the gates) are reported in Table~\ref{exp}. From the table, we can
expect our verification method  should work for quantum circuits of
practical size.  

\begin{table}[tb]
 \begin{center}
\caption{Experimental result} \label{exp}
 \begin{tabular}{|c|c|c|c|} \hline
$\sharp$ variables & $\sharp$ gates  & $\sharp$ nodes & time (sec.) \\ \hline
30 & 100 &  418 & 0.0050   \\ \hline
30 & 200 & 2509 & 0.035  \\ \hline
30 & 400 & 16681 & 16.23 \\ \hline
60 & 100 & 1568 & 0.017  \\ \hline
60 & 200 & 12984 & 10.7664  \\ \hline
60 & 400 & 24681 & 99.58 \\ \hline
\end{tabular}
\end{center}
\end{table}

\section{Conclusions and Future Work}

In this paper, we introduced new concepts: SCQCs together with DDMFs. 
As described, they should be useful for the analysis and the 
verification of quantum circuits with a practical restriction. 
It should be noted that DDMFs are provably useful even for quantum 
circuit design methods since DDMFs can be considered as a  
generalization of the data structure used in the design method in~\cite{AP06}.  

We also revealed the essential difference between DDMFs and QuIDDs
for representing the functionalities of SCQCs. From our comparison, we
can conclude that our approach is  much more efficient for the verification
of SCQCs than a method based on known techniques. 
Note that this does not mean that DDMFs are better than QuIDDs:  
DDMFs are only for SCQCs, whereas QuIDDs can treat all kinds of 
quantum circuits. In other words, in some sense, our approach stands 
in the middle of classical Boolean functions (BDDs) and general quantum
circuit specifications (QuIDDs or QMDDs). As described, this standpoint
can be considered as a good trade-off point if we consider designing 
and analyzing quantum circuits from the practical view point, i.e., when
we focus on sub-circuits to calculate Boolean functions for quantum
algorithms. Lastly we would like to add one more issue: Since DDMFs are 
edge-valued decision diagrams, it may be easier to verify 
{\it quantum phase-equivalence checking} of SCQCs by DDMFs than the method 
based on QuIDDs~\cite{VMH07}. Thus, we consider that it is an
interesting future work to study how efficiently DDMFs 
work for {\it quantum phase-equivalence checking} of SCQCs. 

In conclusion, we can expect the introduction of SCQCs and DDMFs would  
promote research toward practical and  efficient quantum circuit
design methodologies.

\end{document}